\documentclass[final,5p,times,twocolumn,letterpaper]{elsarticle}

\usepackage{natbib}
\usepackage{cancel}
\usepackage{pifont} 
\usepackage{amsthm}         
\usepackage{amsmath} 	   
\usepackage{amssymb}       
\usepackage{amsfonts}
\usepackage{graphicx} 	    
\usepackage{verbatim}
\usepackage{epsfig,bbm}
\usepackage{color}
\usepackage{colortbl}
\usepackage{float}
\usepackage{multirow}

\usepackage{subcaption}

\definecolor{babyblue}{rgb}{0.54, 0.81, 0.94}
\definecolor{corn}{rgb}{0.98, 0.93, 0.36}

\begin{document}

\begin{frontmatter}

\title{Supersmoothing through Slow Contraction}

\author[add2]{William G. Cook}
\author[add0]{Iryna A. Glushchenko}
\author[add1]{Anna Ijjas }
\author[add0]{Frans Pretorius}
\author[add0]{Paul J. Steinhardt}

\address[add2]{Theoretisch-Physikalisches Institut, Friedrich-Schiller-Universit\"at, 07743 Jena, Germany}
\address[add0]{Department of Physics, Princeton University, Princeton, NJ 08544, USA}
\address[add1]{Max Planck Institute for Gravitational Physics (Albert Einstein Institute), 30167 Hanover, Germany}

\date{\today}

\begin{abstract}
Performing a fully non-perturbative analysis using the tools of numerical general relativity, we demonstrate that a period of slow contraction is a ``supersmoothing'' cosmological phase that homogenizes, isotropizes and flattens the universe both classically and quantum mechanically and can do so far more robustly and rapidly than had been realized in earlier studies.     
 \end{abstract}


\end{frontmatter}

{\it Introduction.} 
The degree of homogeneity, isotropy and flatness observed in the universe on large scales is generally viewed as so striking that it calls for some kind of cosmological phase to explain it.  Typically, this very same phase is also supposed to be  the source of a nearly scale-invariant spectrum of  quantum fluctuations spanning length scales much larger than the Hubble radius that, through one means or another, lead to a spectrum of density perturbations.  

To achieve these objectives,  the phase must  be a {\it supersmoother},  meaning it must be a
\begin{itemize}
\item[($i$)]  {\it classical smoother} (the relative contribution of small inhomogeneities and anisotropies to the total energy density must shrink  according to classical cosmological evolution equations); 
\item[($ii$)]
  {\it quantum smoother} (homogenizes and isotropizes even when all quantum fluctuations are included); 
  \item[($iii$)] {\it robust smoother} (insensitive to initial conditions even when they correspond to large, non-perturbative deviations from a homogeneous and isotropic spacetime); and, 
  \item[($iv$)]  {\it rapid smoother} (sufficient smoothing is achieved well before the phase ends).
  \end{itemize}
The goal of this paper is to show that a slow contraction phase \cite{Ijjas:2018qbo} (also known as an ekpyrotic contraction phase \cite{Khoury:2001wf,Buchbinder:2007tw,Levy:2015awa}) satisfies these four conditions.  In fact, it is the only currently known example of a supersmoothing cosmological phase. 

Slow contraction is the mechanism commonly invoked in bouncing and cyclic cosmologies \cite{Ijjas:2018qbo,Ijjas:2019pyf}.   A standard example, and one that will be used here, is described by a canonical scalar field $\phi$ minimally-coupled to Einstein gravity with a negative exponential potential  
\begin{equation}
V(\phi) = -V_0 \,{\rm e}^{- \sqrt{2 \varepsilon}  \phi } \equiv  -V_0 \,{\rm e} ^{ -\phi/M},
\end{equation} 
where $V_0>0$ and  $M^{-1}=\sqrt{2\varepsilon} > 1$. 
(Here and throughout units in which the reduced Planck mass is set equal to one are used.)  

Slow contraction is an attractor scaling solution in which the scalar field homogeneously evolves down the potential as $\phi(t) = \sqrt{2/\varepsilon} \, {\rm ln}\, |t|$ and the scale factor decreases as $a(t) \propto |t|^{1/\varepsilon}$ \cite{Erickson:2003zm}.   In this limit, the parameter $\varepsilon$ characterizes the equation of state; more exactly, $\varepsilon= (3/2)(1+ p/\varrho)$ where $p$ is the pressure and $\varrho$ is the energy density associated with the scalar field. (The contributions of matter, radiation, gradient energy, and all other forms of energy are negligible during the attractor phase.) The slow contraction phase is followed by a classical (non-singular) bounce and reheating ({\it e.g.}, decay of the scalar field energy) to a hot expanding phase with all the large-scale properties of the universe already set as needed to explain cosmological observations.  

Appealing features of bouncing cosmologies of this type are that they are geodesically complete; resolve the cosmic singularity problem; avoid quantum runaway effects that lead to a multiverse of outcomes; and can generate a nearly scale-invariant spectrum of nearly gaussian density perturbations without producing a corresponding spectrum of primary tensor perturbations (in accord with current observations) \cite{Ijjas:2018qbo}. In addition, recently proposed self-similar cyclic versions \cite{Ijjas:2019pyf} avoid the Tolman entropy problem of earlier cyclic cosmologies; enable information to pass smoothly across each bounce; predict the instability of the current vacuum;  and assign dark energy a new role as the critical component shaping the overall cyclic history of the universe.   However, all of these features rely on slow contraction being a supersmoothing phase.  

We begin by reviewing the already-established case that slow contraction is ({\it i}) a classical and ({\it ii}) a quantum smoother.  These conditions can be verified using perturbative analyses around homogeneous spacetimes.  We then turn to new results based on fully non-perturbative calculations utilizing the tools of numerical general relativity that establish ({\it iii}) the robust insensitivity to initial conditions and  ({\it iv}) the rapidity with which slow contraction smooths the universe.    

This investigation builds on early work by Garfinkle et al. \cite{Garfinkle:2008ei} that adapted numerical relativity methods  to study the robustness of slow contraction as a classical smoother beginning from highly non-linear, non-perturbative deviations from homogeneity and isotropy.  Based on case studies, the Garfinkle et al. results suggested that, generically, smoothing never completes.  Rather, slow contraction suffices to make most of the volume homogeneous and isotropic, but there always remains a small regime that is inhomogeneous and anisotropic.  

Here we demonstrate that the case studies of Ref.~\cite{Garfinkle:2008ei} were anomalous in that they inadvertently began with cosmologically implausible initial conditions and limited values of $\varepsilon$.  In our study using still highly non-perturbative but now physically plausible initial conditions, we find that slow contraction for sufficiently large $\varepsilon$ generically results in an entirely smoothed universe and no remnants of inhomogeneity and anisotropy. We further show that, for somewhat larger values of $\varepsilon$,
 the smoothing  completes rapidly after only a tiny amount of contraction, as required to generate a sufficiently broadband spectrum of nearly scale-invariant density perturbations to explain the temperature variations in the cosmic microwave background.
In other words, slow contraction is a significantly more effective and robust smoothing mechanism than suggested by the earlier study.   

\vspace{0.1in}
\noindent
{\it (i) Classical smoother.} We define a phase as a `classical smoother' if the dynamical attractor solution is a flat, homogeneous, and isotropic universe  presuming initial conditions that can be described as small perturbations about a Friedmann-Robertson-Walker (FRW) spacetime and the generalized Friedmann equation,
\begin{equation} 
\label{Fried}
H^2 =   \frac{1}{3} \left( \frac{\rho_m^0}{a^3} +\frac{\rho_r^0}{a^4} +\frac{\rho_{\phi}^0}{a^{2 \varepsilon}}     \right) - \frac{k}{a^2} +\frac{\sigma^2}{a^6}, 
\end{equation} 
where $H \equiv \dot{a}/{a}$ is the Hubble parameter; dot denotes differentiation with respect to the physical time coordinate $t$; the scale factor $a(t)$ is normalized so that $a(t_0)=1$ at $t=t_0$; $\rho_i^0$  represents the  energy density for component $i$ at $t=t_0$; and $i\in\{m,r,{\phi}\}$ refers to the densities of matter, radiation and the scalar field (kinetic plus potential energy density), respectively.  The last two terms correspond to the spatial curvature and anisotropy.  Note that the scalar field gradient energy density in this perturbative limit scales as $1/a^2$,  the same way as the spatial curvature.  

Slow contraction with $\varepsilon>3$ is a classical smoother because the scalar field energy density $\rho_{\phi}^0/a^{2 \varepsilon}$ increases faster than all other terms as $a(t)$ decreases.  Analogously, inflationary expansion \cite{Guth:1980zm,Linde:1981mu,Albrecht:1982wi} with $\varepsilon<1$ passes this test because the scalar field energy density for this equation of state decreases slower than all other terms as $a(t)$ increases.  

A reasonable objection to this test is that it assumes near homogeneity as an initial condition, the same condition that slow contraction (or inflation) is supposed to explain.  The test further assumes that quantum fluctuations  have a negligible effect on the background evolution, which is inconsistent with the fact that inflation is generally eternal \cite{Steinhardt:1982kg,Vilenkin:1983xq} due to large quantum backreaction effects.   That is why the next conditions must also be satisfied for a cosmological phase to be considered as supersmoother mechanism capable of explaining the large-scale properties of the universe.

\vspace{0.1in}
\noindent
{\it (ii.) Quantum smoother.}  A litmus test for a `quantum smoother' is that an initially homogeneous, flat, and anisotropic universe should be stable to quantum fluctuations generated during the smoothing phase.  

A classical smoother is not necessarily a quantum smoother.  In an inflationary phase, for example, quantum fluctuations of the inflaton generate growing mode curvature fluctuations that drive the universe away from homogeneity.   (That there are growing modes traces back in the Mukhanov-Sasaki perturbation equation \cite{Sasaki:1983kd,Kodama:1985bj,Mukhanov:1988jd}  to the fact that $a''/a >0$ for an expanding phase with $\varepsilon<1$, where prime represents $d/d\tau$ and $\tau$ is the conformal time.)

The standard approach in inflationary model-building is to suppress the unstable growth for a range of inflaton field values (corresponding to the last 60 $e$-folds of inflation, say)  by setting  the inflaton self-interaction strength to be exponentially small \cite{Bardeen:1983qw,Steinhardt:1984jj} and setting the initial field strength and kinetic energy density to lie within a specific restricted range.  This is the source of the ``fine-tuning'' and ``initial condition" problems of inflation.  However, there is generally no physical mechanism that can restrain a quantum field like the inflaton from exploring values and kinetic energy densities that lie far outside the chosen restricted range,  including a range of values sometimes called the `self-reproduction' regime.  In regions of space where the field lies in this regime, the quantum-induced perturbations dominate over classical evolution \cite{Steinhardt:1982kg,Vilenkin:1983xq,Linde:1986fd}  and excite growing mode curvature perturbations \cite{Guth:2007ng}.   
The result is a quantum runaway effect in which quantum fluctuations superimposed on quantum fluctuations transform a universe -- {\it even if it is perfectly homogeneous and isotropic universe initially} -- into a spacetime with arbitrary and unpredictable deviations from homogeneity and isotropy.   By definition, this means that inflation fails the litmus test for a quantum smoother, even though it is a classical smoother. 

Failing this test is critical.  It means that, in a fundamental sense, inflation cannot explain the homogeneity (isotropy or spatial flatness) of the universe.   While it is possible that some regions of spacetime are smooth, they are not generic.   

In contrast, the curvature modes decay during slow contraction; {\it e.g.,} the sign of $a''/a$ in the Mukhanov-Sasaki equation \cite{Sasaki:1983kd,Kodama:1985bj,Mukhanov:1988jd} is negative during a contracting phase with $\varepsilon>2$.  Consequently, homogeneity, isotropy and spatial flatness are preserved even when quantum fluctuations are included.  Slow contraction therefore passes the litmus test, meaning that it can actually explain the observed large-scale properties of the universe.  At present, slow contraction is the only known example of a cosmological phase that is both a quantum smoother {\it and} a classical smoother.  

\begin{figure*}[t]
\begin{center}
\includegraphics[width=6.00in,angle=-0]{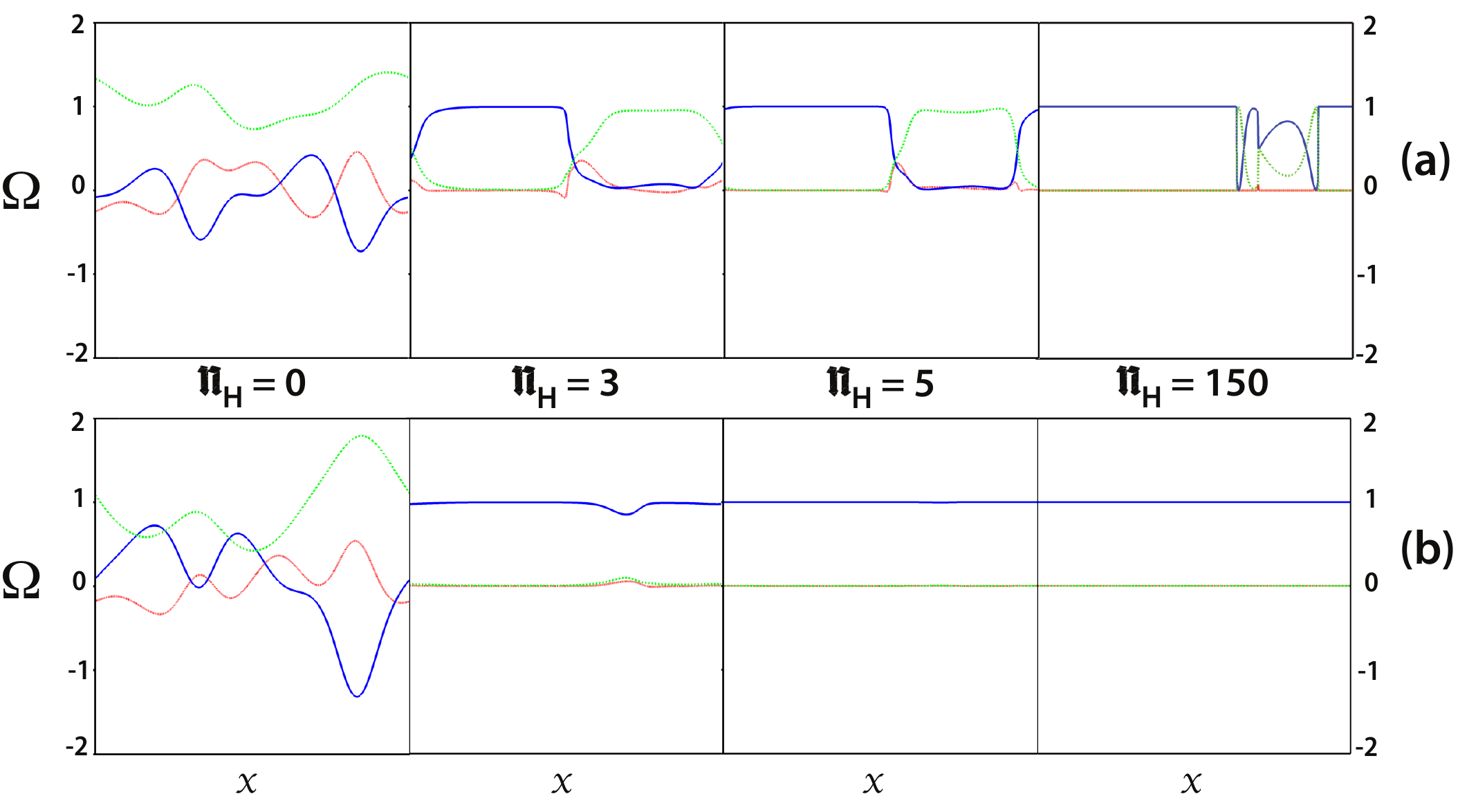}
\end{center}
\caption{Four $t={\rm const.}$ snapshots of the normalized energy density 
in matter $\Omega_m$
(blue, solid line), curvature $\Omega_k$ (red, hashed line) and
shear $\Omega_s$ (green, dotted line) 
for $0\le x \le 2 \pi$ at several times during the evolution for two cases: (a) initial conditions as in Ref.~\cite{Garfinkle:2008ei} with $Q_0=0$  and $f_1$ sufficiently large that the initial scalar field velocity in half the space is wrong-way -- headed very rapidly up the steeply downward potential ($Q<0$); and (b) the same case but with $Q_0=1.5$, sufficiently large that the initial scalar field velocity is headed right-way (downwards) or very slowly wrong-way for all $x$.  $\mathfrak{N}_H$ is the number of e-folds of contraction in the Hubble radius, $|H|^{-1}$. Notably, smoothing is incomplete in case (a) even after $\mathfrak{N}_H =150$ e-folds of contraction as there remains an inhomogeneous region (near $x = 3 \pi /2$ in this example) where the scalar field velocity was initially headed rapidly in the wrong direction; but smoothing completes everywhere if the initial scalar field velocity is nowhere headed rapidly in the wrong direction, even though the initial conditions are highly perturbed with respect to the attractor slow contraction FRW solution, as illustrated in case (b). 
As in Ref.~\cite{Garfinkle:2008ei}, both examples use a physically plausible value of $M=0.1$ (in reduced Planck units), corresponding to $\varepsilon=50$.   
\label{fig1}}
\end{figure*}

\vspace{0.1in}
\noindent
{\it (iii.) Robust smoother.}  A cosmological phase that classically smooths {\it only} for small perturbations away from FRW does not explain the observed homogeneity and isotropy because it fails to smooth for {\it generic} initial conditions.  Hence, a third condition for supersmoothing is that the phase be a `robust' smoother: homogeneity and isotropy should emerge even if the initial conditions are non-perturbatively far away from FRW.   The wider the range of initial conditions that can be smoothed, the more robust is the smoothing phase.  As we shall see, this imposes a somewhat more stringent constraint on the equation of state $\varepsilon$ during a contracting phase than required to be a classical smoother ($\varepsilon>3$) or quantum smoother ($\varepsilon>2$).

To analyze the robustness of slow contraction to a wide range of initial conditions with non-perturbative deviations from FRW,  we solve the full 3+1 dimensional Einstein-scalar field equations numerically beginning from large initial inhomogeneous spatial curvature, matter density and shear and track their evolution for long times ({\it i.e.,} up to several hundreds of $e$-folds).  As in Ref.~\cite{Garfinkle:2008ei}, we restrict ourselves to deviations from homogeneity along a single spatial direction so that the spacetimes have two Killing fields, although with sufficiently general initial data that the behavior approaching the singularity is the same as with no restriction.   The scheme is fully detailed in Ref.~\cite{Paper2}.  (We have extended our numerical simulations to cases where the inhomogeneities are along two dimensions that will be presented elsewhere \cite{Paper4}, but we have not observed any qualitative differences in the result.)   

We note that, at present, there is no analogous test of robustness for an expanding case, including inflation.  Recent work has used full numerical relativity simulations to explore smoothing of large initial inhomogeneities
in inflation~\cite{East:2015ggf,Clough:2016ymm},
though these early studies have only considered initial data where the
scalar field perturbation has spatially uniform velocity which is set to zero. This is a rather special initial condition that favors inflation and is arguably far from what might be expected as a generic pre-inflationary state.

A key feature of our scheme is to use scale-invariant (Hubble-normalized) variables (denoted by bar). For example, we define the scalar field time derivative
\begin{equation}
\bar{W} = {{\cal N}^{-1}} {\partial _t} \phi 
\label{wdef}
\end{equation}
where ${\cal N} = N/\Theta$ is the scale-invariant generalization of the lapse $N$, and
the time coordinate $t$ is given through
\begin{equation}
e^t = {\textstyle \frac13} \Theta,
\label{timechoice}
\end{equation}
with $\Theta = | H^{-1} |$, so that surfaces of constant time are constant mean curvature hypersurfaces. Note that, in the homogeneous limit, the variables of our numerical scheme reduce to the well-known dimensionless Friedmann variables, $\Omega_i$ for component $i$, representing the fractional contribution of component $i$ (matter density, curvature or anisotropy) to  $H^2$ in the Friedmann equation, see {\it e.g.} \cite{Levy:2015awa}.  Note that the matter contribution ($\Omega_m$), which includes the sum of positive kinetic energy density and negative potential energy density, and the curvature contribution  ($\Omega_k$) can be positive or negative.

We set the initial conditions by first picking a particular time $t_0$. 
Then, for the geometry we must provide the spatial metric as well as the
extrinsic curvature of the $t_0$-hypersurface. Not all components of these tensors
are freely specifiable, but must satisfy the constraint equations of general relativity.
(Notably, the evolution equations propagate the constraints, {\it i.e.,} ensure that the constraints are satisfied at later times.)
To this end, we adapt the York method \cite{York:1971hw} commonly used in numerical relativity computations:  We freely specify a conformally flat initial metric and the vacuum contribution 
to the conformally rescaled trace-free extrinsic curvature.
This method enables us to freely choose the initial field $\phi(x,t_0)$ and velocity distributions 
\begin{equation}
\bar{W}(x,t_0) \equiv \psi ^{-6}(x,t_0) Q(x,t_0),
\end{equation}
as well as the divergence-free part of the initial shear contribution, which is the trace-free part of the extrinsic curvature  
\begin{equation}
\bar{\Sigma} _{ab}(x,t_0) \equiv \psi ^{-6}(x,t_0) Z_{ab}(x,t_0).
\end{equation}
The set of initial data is completed by solving the constraint equations for the conformal factor $\psi(x,t_0)$ and the rest of $Z_{ab}(x,t_0)$.

Periodic boundary conditions with $0 \le x \le 2 \pi$ with $0$ and $2 \pi$ 
identified are used; hence, functions $x$ can be expressed as sums of Fourier modes.  
We use the same divergence-free and trace-free ansatz for 
$Z_{ab}$ as in Ref.~\cite{Garfinkle:2008ei}.

A critical difference  from Ref.~\cite{Garfinkle:2008ei} is the inclusion of a homogeneous term, $Q_0$, in the initial conditions for $Q$,
\begin{equation}
\label{Qic}
Q(x,t_0) = \Theta \, (f_1 \cos(m_1 x + d_1)  + Q_0),
\end{equation}
where $Q_0,f_1, m_1$ and $d_1$  are constants.  The test for robustness entails highly non-linear initial conditions such that $\Delta_f= f_1/Q_{attr}$ and $\Delta_Q= |Q_0-Q_{attr}|/Q_{attr}$ are ${\cal O}(1)$, where $Q(x,t) = Q_{attr}(t)$ is the homogeneous attractor solution.

$Q_0$  represents the average value of $Q$ over the periodic box.  Including only the cosine perturbation term proportional to $f_1$, as was done in Ref.~\cite{Garfinkle:2008ei}, means that, for a portion of the space, the initial scalar field velocity is aimed up the steep potential rather than down. Revisiting the analysis in Ref.~\cite{Garfinkle:2008ei}, we find that precisely the regions with the maximal ``wrong-way'' initial velocity end up being trapped in inhomogeneous regions (see Fig.~1a) where the scalar field changes on very small spatial scales and behaves like a fluid with $w=1$ (that is, as if there were no potential) and where the dynamical behavior is similar to chaotic mixmaster vacuum solutions.   (Note that our sign convention is that positive $Q$ corresponds to rolling the right way, {\it i.e.,} downhill, and negative $Q$ corresponds to the wrong-way.)

In the context of bouncing cosmology, though, the conditions in which the scalar field is initially rolling rapidly up a steeply downward potential  at the beginning of the contracting phase are quite extreme and, in some contexts, physically nonsensical.
For example, in a cyclic cosmology, the transition from a slowly accelerated expansion phase (like the current dark energy dominated phase) to a slowly contracting phase occurs only if the scalar field first rolls {\it down the potential} (right-way) sufficiently far such that the potential energy density changes from slightly positive to sufficiently negative.  In other words, rolling down the steep potential is a prerequisite for contraction to begin and, therefore, the right-way condition is automatically satisfied.   The study in Ref.~\cite{Garfinkle:2008ei} did not consider this physical requirement, and, as illustrated by Fig.~1b,  this is the {\it only} reason why the inhomogeneous region formed during slow contraction (as shown in Fig.~1a).  
For the case illustrated in Fig.~1b, 
  $Q_0$ has been set to a sufficiently positive value that the initial scalar field velocity is not rapidly uphill (negative) for any $x$, as expected physically, and   the evolution converges to a smooth solution despite the fact that the initial conditions are highly non-perturbative  in the sense that $\Delta_f$ and $\Delta_Q = {\cal O}(1)$, the same as in Ref.~\cite{Garfinkle:2008ei}.

\begin{figure}[t]
\begin{center}
\includegraphics[width=2.5in]{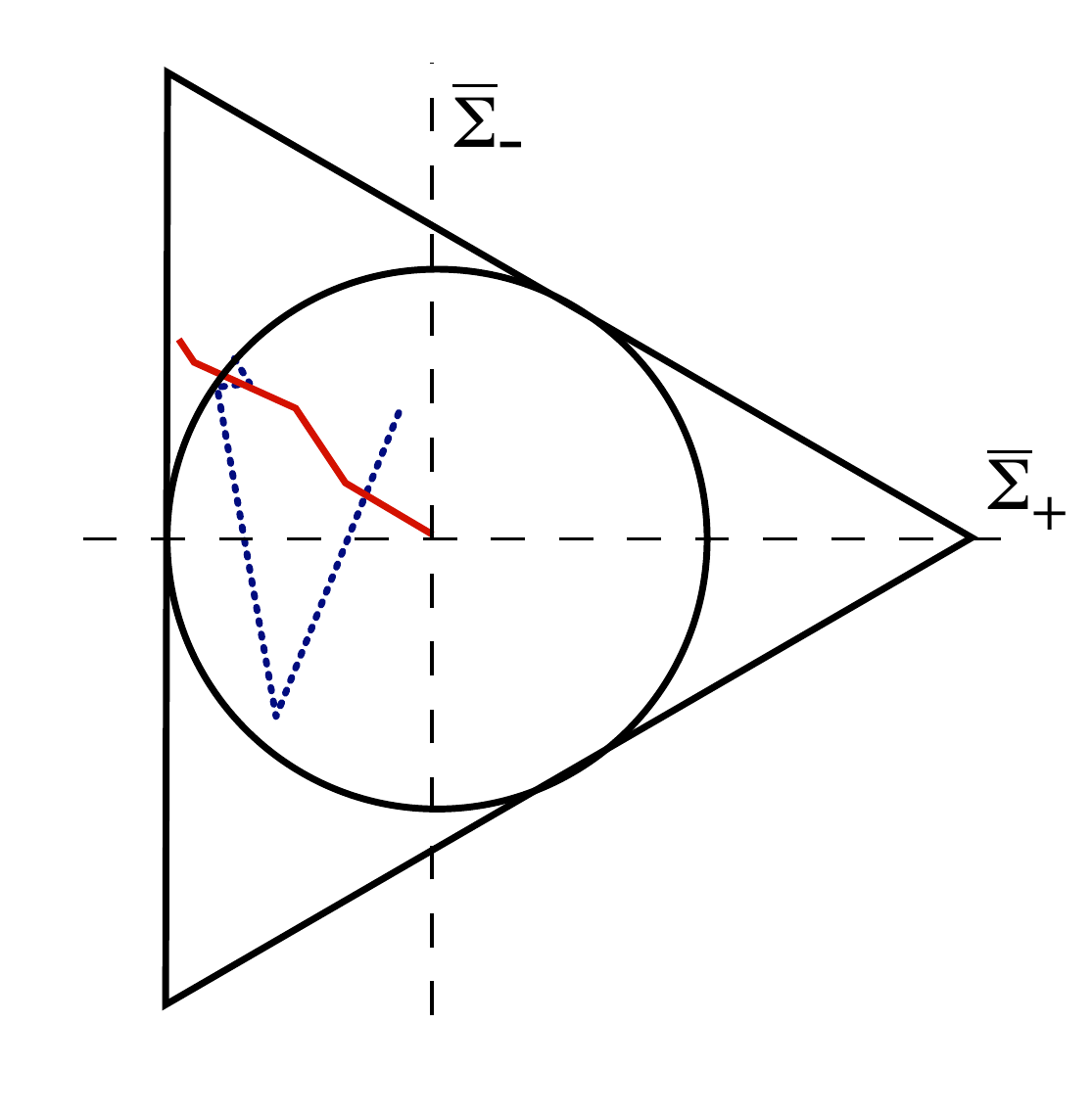}
\end{center}
\caption{The state space orbit for a worldline at
$x=3\pi/2$, a point in the inhomogeneous region for the case with $Q_0=0$, as shown in Fig. 1a (blue, dotted); and then superposed the space orbit for the same point the case with $Q_0>0$ shown in Fig.~1b (red, solid).  The center of the circle corresponds to an isotropic FRW universe.  The first  orbit (blue, dotted) corresponding to an inhomogeneous region
 never reaches the center,  whereas the second example (red, solid) corresponding to a smoothed region does reach the center.  
\label{fig2}}
\end{figure}

Fig.~\ref{fig2} shows the state space orbits associated with Figs.~1a and~1b evaluated at the same value of $x =3 \pi/2$ projected onto 
the
$(\bar{\Sigma}_+ ,\bar{\Sigma}_-)$ plane, where
\begin{equation}
\bar{\Sigma}_+ = \textstyle{\frac12}\Big(\bar{\Sigma}_{11} + \bar{\Sigma}_{22}\Big)
,\quad
\bar{\Sigma}_- = \textstyle{\frac{1}{2\sqrt{3}}}\Big(\bar{\Sigma}_{11} - \bar{\Sigma}_{22}\Big).
\end{equation}
The point $x =3 \pi/2$ lies in the inhomogeneous region of Fig.~1a.  The orbits begin near the outer (Kasner) circle and travel inward.  They show that, for the case in Fig.~1a, the orbit never converges to the center (corresponding to FRW), signifying that the mixmaster-like reflections in the inhomogeneous region never isotropize.  This is to be contrasted with the orbit shown for the case in Fig.~2b tracking the same point $x =3 \pi/2$ which isotropizes and converges to the center of the plot.

\begin{figure}[t]
\begin{center}
\includegraphics[width=2.70in]{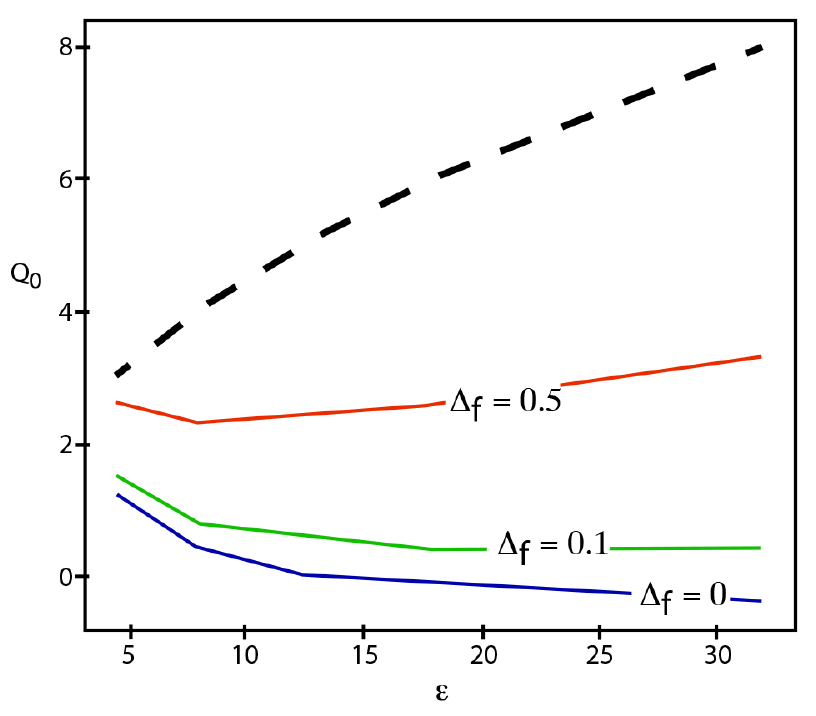}
\end{center}
\caption{The three solid curves show the minimal value of $Q_0$ required for complete smoothing and convergence to the attractor solution $Q(x,t)=Q_{attr}$ (bold dashed curve) as a function of $\varepsilon$, where  $\Delta_f= f_1/Q_{attr}$ measures the spatial inhomogeneity of the initial velocity distribution.  (The initial distribution of $Z_{ab}$ is homogeneous.)   The solid curve marked $\Delta_f= 0$ corresponds to a strictly homogeneous initial condition and the upper solid curves correspond to increasingly inhomogeneous initial velocity distributions.      
\label{fig3}}
\end{figure}

The critical value of $Q_0$ required to have ``complete smoothing'' (smoothing for all $x$) depends on $\varepsilon$ and $\Delta_f= f_1/Q_{attr}$.  The curves in Fig.~\ref{fig3} show the critical $Q_0$ as a function of $\varepsilon$ for three different values of $\Delta_f$ for cases in which the initial $Z_{ab}$ distribution is homogeneous. 
Depending on $\Delta_f$, complete smoothing and convergence to the attractor solution $Q(x,t)=Q_{attr}(t)$ occurs for any value of $Q_0$ on or above the corresponding curve.  
Note that there is a significant gap between these curves and the curve corresponding to the attractor solution, $Q(x,t)=Q_{attr}(t)$, indicating that the initial velocity may be quite far from the eventual smooth solution; this is a sign of robustness.  
 
 The larger the value of $\varepsilon$ is, the greater the robustness is.  For example, for the case of a homogeneous initial velocity distribution (the curve marked $\Delta _f=0$), complete smoothing  and convergence to the attractor solution occurs for $\varepsilon \gtrsim 13$ (or $M^{-1}>5.1$) even if the scalar field begins at rest for all $x$ ($Q_0=0$). For cases with non-uniform initial velocity ($\Delta_f \ne 0$), it suffices if 
  $Q_0$ is positive enough that the initial scalar field velocity is downhill for all $x$.   As noted above, rolling downhill is typically a prerequisite for contraction to begin (and absolutely necessary for cyclic models) and so, in these scenarios, this condition is automatically satisfied. 

%

\vspace{0.1in}
\noindent
{\it (iv.) Rapid smoother.}  A measure of smoothing power is how rapidly a cosmological phase can transform a highly inhomogeneous and anisotropic patch into a nearly FRW universe like the one we observe.  The rapidity is critically important in cases like bouncing cosmology or inflation where the same phase is supposed to generate a nearly scale-invariant spectrum of fluctuations on scales larger than the Hubble radius during the smoothing period \cite{Levy:2015awa}.  In these cases, the smoothing phase must last long enough to first homogenize the background beginning from highly non-perturbative initial conditions and, {\it  in addition,} last long enough to generate the requisite band of quantum fluctuations on the smoothed background as needed to explain the cosmic microwave background and galaxy formation.  

The case of cyclic bouncing cosmology \cite{Ijjas:2019pyf} is the most restrictive because the slow contraction phase is limited to a finite period beginning when the Hubble radius is roughly the current value $|H_{beg}|^{-1} = {\cal O}(10^{28}$~cm) and ending  when the Hubble radius shrinks to a microscopic size $|H_{end}|^{-1} = {\cal O}(10^{-25}$~cm).  There follows a classical (non-singular) bounce to an expanding phase accompanied by reheating.  Note that $|H_{end}|^{-1}$   is much larger than the Planck length where quantum gravity effects are non-negligible.  This value of $|H_{end}|^{-1}$  corresponds to a reheat temperature of $T\sim10^{15}$~GeV.  In total, the change in the Hubble radius during the slow contraction phase is a factor of about 120 $e$-folds. 

This range of 120 $e$-folds determines the maximum range of wavelengths (Fourier modes) that exit the Hubble radius during the slow contraction phase.  During slow contraction, fluctuations exit the Hubble radius because the Hubble radius shrinks rapidly while the scale factor $a(t)$ changes negligibly (see below).  Each mode can be labeled by wavenumber $k/a = H_{exit}$, that depends on the value of the Hubble radius when the mode exits the Hubble radius.      Consequently,
quantum fluctuations with wavelengths spanning the roughly 120 e-folds
between $k/a = |H_{beg}|^{-1}$ and $k/a = |H_{end}|^{-1}$ 
exit the Hubble radius by the time the bounce is reached.   

After the bounce, $|H|^{-1}$ expands in proportion to $a^2$ during the radiation-dominated phase and as $a^{3/2}$ during the matter-dominated phase.  The scale factor $a(t)$ grows by about 60 e-folds over this period.  Hence, modes with with comoving wavenumbers  between $k/a = |H_{end}|^{-1}$ and $k/a = e^{60} |H_{end}|^{-1}$  lie within the Hubble radius today.  These are probed by observations of galaxy formation and the temperature fluctuations of the cosmic microwave background radiation.   In order to agree with observations, these fluctuations should be the only deviations from homogeneity and isotropy; that is, the initial highly-nonlinear deviations from FRW at the beginning of slow contraction must be smoothed well  before these last 60 e-folds of quantum fluctuations exit the Hubble radius.  
(The first 60 e-folds to exit the Hubble radius lie beyond the current Hubble radius and are unconstrained by  observations.)   The rapidity constraint, therefore,  is that smooth contraction must be rapid enough to smooth the universe well within the first 60 e-folds of the contraction of $|H|^{-1}$.  

The rapidity depends on the rate of slow contraction which, in turn, depends on the equation of state $\varepsilon$. We have seen that the minimal value required for a classical and quantum smoother is $\varepsilon=3$.   We have shown in Figs.~2 and~3 above that,  for $M^{-1}\gtrsim 5.1$ or $\varepsilon \gtrsim 13$, the smoothing is robust: the universe is completely smoothed for the entire range of physically plausible initial conditions.  

Now we want to consider how long the smoothing takes.  For $4 \lesssim M^{-1} \lesssim 10$, the smoothing is complete within the first 60 $e$-folds or less, depending on how nonlinear the initial conditions are.  This barely meets the minimum criterion for rapidity.   For modestly greater values, $M^{-1} \gtrsim 10$ or $\varepsilon\gtrsim 50$, corresponding to our example in Fig.~1b above, the smoothing is complete in less than 10 $e$-folds, easily satisfying the rapid smoother condition.  

\vspace{0.1in}
\noindent
{\it Discussion.}   Explaining the observed large-scale properties of the universe requires a mechanism that causes these properties to emerge even from initial conditions that are very different from the desired outcome.  
We have argued that supersmoothing, as defined by the four criteria described above, is required to achieve this goal. Then, using the tools of numerical relativity, we have shown that a slow contraction phase with  $\varepsilon \ge {\cal O}(10)$ satisfies these criteria,  the only example of a supersmoother phase currently known.  

An additional notable property of the slow contraction phase is that the scale factor $a(t)$ hardly shrinks at all  (by only $\mathfrak{N}_a = \ln a/a_{beg} \approx \mathfrak{N}_H/\varepsilon$  $e$-folds) compared to the Hubble radius (which shrinks by $\mathfrak{N}_H$  $e$-folds).  
  For example, in the case shown in Fig.~1b,  $a(t)$ shrinks by  $\mathfrak{N}_a=2.2$ e-folds during the same period that the Hubble radius shrinks by $\mathfrak{N}_H \approx 150$ $e$-folds.   In a cyclic bouncing model, this means that the distance between black holes (or galaxies) existing at the end of an expanding phase decreases by a negligible amount  over the entire contraction phase! 

Consequently, there is no crunch of pre-existing black holes, galaxies or other macroscopic objects as the bounce approaches. There is also no large increase in the density of ordinary matter and radiation.  Space remains classical and spread out.   What changes exponentially is the size of the Hubble radius and the energy density stored in the scalar field driving slow contraction.  This is a fundamental difference between cyclic models based on slow contraction and classical non-singular bounces versus all previous cyclic models dating back to Friedmann and Tolman or early conceptions envisioned before the introduction of general relativity \cite{Ijjas:2019pyf}.  The entropy problem that plagued earlier renditions is not relevant here because nearly all entropy from the previous cycles lies outside the Hubble radius at the bounce and does not re-enter after the bounce.

 \vspace{0.1in}
\noindent
{\it Acknowledgements.} 
W.G.C. is partially supported by the Simons Foundation grant number 654561. The work of A.I. is supported by the Lise Meitner Excellence Program of the Max Planck Society and by the Simons Foundation grant number 663083.
F.P. acknowledges support from NSF grant PHY- 1912171, the Simons Foundation, and the Canadian Institute For Advanced Research (CIFAR).  P.J.S. is supported in part by the DOE grant number DEFG02-91ER40671 and by the Simons Foundation grant number 654561.  A.I. thanks the Black Hole Initiative at Harvard University for hospitality, where parts of this work were completed.

\bibliographystyle{apsrev}
\bibliography{Q0}

\end{document}